\documentstyle[11pt]{article}
\begin{document}

\begin{center}
{\LARGE \bf Quasi-uniform approximation for the radial
Schr\"odinger equation with power-law potentials}\\
\medskip
{\bf V. V. Kudryashov and Yu. V. Vanne}\\
\medskip
{B.I.Stepanov Institute of Physics, National Academy
of Sciences of Belarus}\\

{F.Skaryna avenue 68, 220072 Minsk, Belarus} \\

E-mail: kudryash@dragon.bas-net.by

\end{center}

\begin{abstract}{The uniformly valid approximation to solutions of the
radial Schr\"odinger equation with power-law potentials are
obtained by means of the explicit summation of the leading
constituent WKB series. }
\end{abstract}



\medskip

Similar questions arise when we use the perturbation theory or the
variational method in order to solve the radial Schr\"odinger
equation
\begin{equation}
\frac{d^2 \Psi(r)}{d r^2} =  \frac{Q(r)}{\hbar^2} \Psi(r) , \quad
Q(r) = 2m \left( V(r) +\frac{\hbar^2 l (l+1)}{2 m r^2} - E\right)
 .
\end{equation}
How to find the unperturbed Hamiltonian or how to find the trial
function for an arbitrary given potential? The universal answers
are absent. In this sense both mentioned methods are not
complete.  At the same time the WKB approximation is directly
determined by a given potential. The WKB approach deals with a
logarithmic derivative $$ Y(r) =\frac{d \ln \Psi(r)}{dr}, \quad
\Psi(r) = \exp\left(\int^rY(r')dr'\right) $$ which satisfies the
nonlinear Riccati equation
\begin{equation}
\frac{dY(r)}{dr} + \left(Y(r)\right)^2 = \frac{Q(r)}{\hbar^2} .
\end{equation}
In this approach two independent solutions $ Y^{\pm}(r) $ of the
Riccati equation are represented by their asymptotic expansions
\begin{equation}
Y^{\pm}_{as}(r) =  \hbar^{-1} \left(\pm  Q^{1/2} +
 \sum_{n = 1}^{\infty}\hbar^n Y^{\pm}_n (r) \right)
\end{equation}
in powers of Plank's constant $\hbar$. The usual WKB
approximation contains a finite number of leading terms
$Y_n^{\pm}(r)$ from the complete expansion $ Y^{\pm}_{as}(r)$.
The conventional WKB approximation is not valid at turning points.

 As it is well known, the WKB
series is divergent. The direct summation of a divergent series
does not exist. By summing one means finding a function to which
this series is the asymptotic expansion \cite{bend1}. There are
several investigations on properties of the WKB terms \cite{bend2,
robn}. In recent years many studies have been devoted to
extracting some useful information about the exact eigenfunctions
from the divergent WKB series (see for instance \cite{pham} and
references therein).
 In our previous work \cite{kudr1} we reconstructed the
WKB series as a sum of new constituent (partial) asymptotic
series. The explicit summation of the leading constituent series
was performed. As a result the new quasi-uniform approximation to
logarithmic derivatives was obtained. This approximation was
derived formerly \cite{kudr2} by other means.  Our approximation
reproduces the known satisfactory approximation \cite{bend1} near
turning points. The proposed method was verified in the case of
the one-dimensional Schr\'odinger equation without singularities
\cite{kudr3}. In  the present work we apply our approach to the
radial Schr\'odinger equation with the centrifugal term which is
singular at the origin ($r=0$).

The second-order quasi-uniform approximation has the following
form
\begin{equation}
Y_{ap}(r ) = Y(r;t) = b_1(r) y_1(a;t) + b_2(r) y_2(a;t)
\end{equation}
where
\begin{equation}
a(r) = \frac{1}{\hbar^{2/3}}\frac{Q(r)}{|Q'(r)|^{2/3}} , \quad
b_1(r) =\frac{1}{\hbar^{2/3}} \frac{Q'}{|Q'|^{2/3}} ,\quad
b_2(r)=  \frac{Q''}{Q'} .
\end{equation}
Here we introduce notations for logarithmic derivatives
\begin{equation}
y_1(a;t) = \frac{d}{da}\ln\left({\rm Ai}(a) + t {\rm Bi}(a)\right)
\end{equation}
of linear combinations of well studied Airy functions {\rm Ai}(a)
and {\rm Bi}(a) \cite{hand}. The value of an arbitrary  parameter
$t$ determines a selected particular solution. We also use
notation
\begin{equation}
y_2(a;t) =  \frac{1}{30} \left[-8 a^2 (y_1(a;t))^2  -4 a y_1(a;t)
+ 8 a^3 - 3 \right] .
\end{equation}

We consider power-law potentials
\begin{equation}
V(r) = \alpha r^k ,\quad \alpha > 0 , \quad k > 0 .
\end{equation}
When $l = 0$ we deal with the usual one-dimensional problem for
odd states which was solved formerly. When $l\neq 0$ we have a
new problem. In this case there are two turning points $r_-$ and
$ r_+$ ($r_+ > r_-$) which satisfy equation $Q(r) = 0$. Turning
points separate three regions.

In the first region where  $r > r_+$ we must describe only the
decreasing solution of the Schr\"odinger equation. Therefore in
this case we select the unique particular solution $Y(r;0)$ of
the Riccati equation.

In the second region where $r_- < r < r_+$ we must describe
oscillating solutions of the Schr\"odinger equation. Therefore in
this case we select two particular solutions $Y(r;\pm i)$ of the
Riccati equation.

In the third region where $r < r_-$ we must reproduce the known
behaviour of the exact solution $ \Psi_{ex}(r) $ of the
Schr\"odinger equation when $ r \rightarrow 0$. In this case we
have following simple formulas
\begin{equation}
Q(r) \rightarrow \frac{\hbar^2 l (l+1)}{r^2} ,\quad \Psi_{ex}(r)
\rightarrow r^{l+1} , \quad Y_{ex}(r) \rightarrow \frac{l+1}{r} .
\end{equation}
At the same time we can derive relations
\begin{equation}
a(r) \rightarrow a_l = \left(\frac{l (l+1)}{4}\right)^{1/3} ,
\quad b_1(r) \rightarrow -\frac{2 a_l}{r} , \quad b_2(r)
\rightarrow -\frac{3}{r} .
\end{equation}
Thus, the approximate logarithmic derivative tends as follow
\begin{equation}
Y(r;t) \rightarrow \frac{c(l,t)}{r}
\end{equation}
where
\begin{equation}
 c(l,t) = - \left(2 a_l y_1(a_l;t) + 3 y_2(a_l;t)\right) .
\end{equation}
Naturally we must demand
\begin{equation}
c(l,t) = l + 1 .
\end{equation}
It is the  quadratic algebraic equation for determining the value
of $t$. One of two solutions leads to an unphysical singularity.
The second solution (nearest to zero ) $t_l$ gives dependence of
$t$ on $l$. We present two numerical examples $t_1 = -0.00393839,
t_2 = -0.000523169$. Hence, in third region we select the unique
particular solution $Y(r;t_l)$ of the Riccati equation.

Matching particular solutions at turning points we obtain the
continuous approximate  wavefunction   $\Psi_{ap}(r)$ which is
represented by following formulas
\begin{equation}
\Psi_1(r) = N \cos{\phi_l}
 \exp\left( -\int_r^{r_-} Y(r';t_l) dr' \right)
\end{equation}
when $r < r_-$,
\begin{eqnarray}
\Psi_2(r) = N \exp \left( \int _{r_-}^r
 \frac{Y(r';+i) + Y(r';-i)}{2} d r' \right) \nonumber
\\ \times
 \cos \left(\int_{r_-}^r \frac{Q'}{|Q'|}
 \frac{Y(r';+i) - Y(r';-i)}{2 i} d r' - \phi_l
 \right)
\end{eqnarray}
when $r_- < r < r_+$ and
\begin{equation}
\Psi_3(r) = \frac{N}{2}(-1)^n
 \exp\left( \int_{r_-}^{r_+} \frac{Y(r';+i) +
 Y(r';-i)}{2} d r'\right)
 \exp\left( \int_{r_+}^r Y(r';0) d r' \right)
\end{equation}
when $r > r_+$. Here
\begin{equation}
\phi_l = \frac{\pi}{3} -  \arctan t_l
\end{equation}
and $N$ is a normalization constant.

  We have the new quantization condition
\begin{equation}
\int_{r_-}^{r_+} \frac{Q'}{|Q'|} \frac{Y(r;+i) -
 Y(r;-i)}{2 i} d r
= \pi (n + \frac{1}{3}) + \phi_l, \quad n = 0,1,2...
\end{equation}
which determines the spectral value  $E_{sp}(n,l)$ of energy
implicitly. We denote the approximate wavefunctions with $E =
E_{sp}(n,l)$ as $\Psi_{ap}(r;n,l)$.

So the approximate eigenfunctions are defined completely. But the
question arises about the optimal approximate eigenvalues because
the value $E_{sp}(n,l)$ is not unique choice.

  Since explicit expressions for wavefunctions are already obtained
we are able to calculate expectation values
\begin{equation}
\bar E(n,l) = <\Psi_{ap}(n,l)|\hat H_r|\Psi_{ap}(n,l)>
\end{equation}
of the  radial Hamiltonian
\begin{equation}
\hat H_r = -\frac{\hbar^2}{2 m}\frac{d^2}{d r^2} + V(r) +
\frac{\hbar^2 l (l+1)}{2 m r^2}
\end{equation}
where $|\Psi_{ap}(n,l)>$ is the vector in the Hilbert space which
corresponds to the function $\Psi_{ap}(r;n,l)$.

  In accordance with the eigenvalue problem
$ \hat H |\Psi> -E |\Psi> = 0$ we construct the discrepancy vector
\begin{equation}
|D(e;n,l)> = \hat H_r |\Psi_{ap}(n,l)> - e |\Psi_{ap}(n,l)>
\end{equation}
with an arbitrary parameter $e$ while $\hat H_r$ and
$|\Psi_{ap}(n,l)>$ are given. It is natural to require that the
discrepancy vector should not contain a component proportional to
the approximate eigenvector. In other words  we consider the
orthogonality condition $<\Psi_{ap}(n,l)|D(e;n,l)> = 0$ as a
criterion for selection of the optimal approximate eigenvalue. As
a result we get just  the expectation value $\bar E(n,l)$ while
$E_{sp}(n,l)$ does not fulfil the above requirement. It should be
also noted that the scalar product $<D(e;n,l)|D(e;n,l)>$ is
minimized at $e = \bar E(n,l)$.

  Now we must verify our approximation numerically for the exactly
solvable problems. We shall compare the normalized approximate
wavefunctions $\Psi_{ap}(r;n,l)$ with the normalized exact
wavefunctions $\Psi_{ex}(r;n,l)$.

   In order to estimate the closeness of two functions $f_1(r)$
and $f_2(r)$ we consider two corresponding vectors $|f_1>$ and
$|f_2>$ in the Hilbert space. Then we construct a deviation vector
$ |\Delta f> = |f_1> - |f_2>$  and a scalar product
\begin{equation}
<\Delta f|\Delta f> = <f_1|f_1> + <f_2|f_2>  - <f_1|f_2> -
<f_2|f_1> .
\end{equation}
Now we can define the relative deviation
\begin{equation}
\delta f = 1 - \frac{<f_1|f_2> + <f_2|f_1>}{<f_1|f_1> + <f_2|f_2>}
\end{equation}
as the numerical estimation of the closeness of two functions.
Note that $\delta f = 0$ when $f_1(r) = f_2(r)$.

   Thus, we get the following estimation
\begin{equation}
\delta\Psi(n,l) = 1 - <\Psi_{ex}(n,l)|\Psi_{ap}(n,l)>
\end{equation}
in the case of normalized real functions $\Psi_{ex}(r;n,l)$ and
$\Psi_{ap}(r;n,l) .$ The same numerical comparison may be
performed
\begin{equation}
\delta\Psi'(n,l) = 1
 - \frac{2 <\Psi'_{ex}(n,l)|\Psi'_{ap}(n,l)>}
{<\Psi'_{ex}(n,l)|\Psi'_{ex}(n,l)> +
<\Psi'_{ap}(n,l)|\Psi'_{ap}(n,l)>}
\end{equation}
for first derivatives $\Psi'(r) = d\Psi(r)/d r .$ Naturally, we
can define analogous estimations for higher derivatives.

     Now we compare two functions $\hat H _r\Psi_{ap}(r;n,l)$
and $\bar E(n,l) \Psi_{ap}(r;n,l)$. This comparison may be
performed when we do not know exact solutions. As a result we get
the relative discrepancy
\begin{equation}
d(n,l) = \frac{<\Psi_{ap}(n,l)|\hat H_r^2|\Psi_{ap}(n,l)> - \bar
E(n,l)^2} {<\Psi_{ap}(n,l)|\hat H_r^2|\Psi_{ap}(n,l)> + \bar
E(n,l)^2}
\end{equation}
which is directly connected with the initial Schr\"odinger
equation under consideration.

 Finally, we characterize our approximation by the usual relative
energy error
\begin{equation}
\delta E(n,l) = \frac{\bar E(n,l)}{E_{ex}(n,l)} - 1
\end{equation}
where $E_{ex}(n,l)$ is the exact energy value.

Firstly, we estimate our approximation in the case of the
oscillator potential $ V(r) = \alpha_2 r^2 $ for which the exact
wavefunctions are well known \cite{flug}.Table 1 shows that the
quasi-uniform approximation gives fairly accurate results for all
considered quantities in the oscillator case.

\begin{center}
Table 1. $ V(r) = \alpha_2 r^2.$

\medskip
\begin{tabular}{|c|c|c|c|c|}\hline
$n,l$ &$\delta\Psi(n,l)$ &$\delta\Psi'(n,l)$ &$d(n,l)$ &$\delta
E(n,l) $ \\ \hline
 0,0  &$1.595\;10^{-5}$ &$7.118\;10^{-5}$ &$6.087\;10^{-5}$ &$5.635\;10^{-5}$ \\
 1,0  &$1.504\;10^{-6}$ &$3.098\;10^{-6}$ &$7.312\;10^{-7}$ &$1.884\;10^{-6}$ \\
 2,0  &$4.271\;10^{-7}$ &$6.391\;10^{-7}$ &$8.413\;10^{-8}$ &$3.348\;10^{-7}$
 \\ \hline
 0,1  &$1.132\;10^{-3}$ &$6.625\;10^{-3}$ &$9.475\;10^{-3}$ &$3.083\;10^{-3}$ \\
 1,1  &$1.337\;10^{-3}$ &$2.898\;10^{-3}$ &$7.719\;10^{-4}$ &$1.011\;10^{-3}$ \\
 2,1  &$1.506\;10^{-3}$ &$2.891\;10^{-3}$ &$5.863\;10^{-4}$ &$7.271\;10^{-4}$
 \\ \hline
 0,2  &$5.052\;10^{-4}$ &$4.877\;10^{-3}$ &$4.627\;10^{-3}$ &$1.492\;10^{-3}$ \\
 1,2  &$4.173\;10^{-4}$ &$8.886\;10^{-4}$ &$1.153\;10^{-4}$ &$2.487\;10^{-4}$ \\
 2,2  &$4.635\;10^{-4}$ &$8.646\;10^{-4}$ &$8.713\;10^{-5}$ &$1.813\;10^{-4}$
 \\ \hline
\end{tabular}
\end{center}

The second example is the linear in $r$ potential $ V(r) =
\alpha_1 r $ . When $l = 0$ the quasi-uniform approximation
reproduces the exact result. When $l \neq 0$ the exact
wavefunctions are unknown but there are results of numerical
solution for  some energies \cite{eich}. We calculate values of
$d$ and $\delta E$ for these states.Table 2 demonstrates validity
of our approximation in the linear case.

\begin{center}
Table 2. $ V(r) = \alpha_1 r$

\medskip
\begin{tabular}{|c|c|c|}\hline
$n,l$  &$d(n,l)$        &$\delta E(n,l) $\\ \hline
 0,1  &$6.951\;10^{-3}$ &$2.684\;10^{-3}$ \\
 1,1  &$5.357\;10^{-4}$ &$8.230\;10^{-4}$ \\ \hline
 0,2  &$2.565\;10^{-3}$ &$1.160\;10^{-3}$ \\
 1,2  &$8.652\;10^{-5}$ &$2.238\;10^{-4}$ \\ \hline
\end{tabular}
\end{center}

 Hence, the performed reconstruction of the
WKB series and subsequent explicit summation of the leading
constituent (partial) series yield the satisfactory (qualitative
and quantitative) description of wavefunctions in the case of the
radial Schr\"odinger equation with power-law potentials.


\end{document}